\RequirePackage{lineno}
\documentclass[preprint,showpacs,preprintnumbers,amsmath,amssymb,superscriptaddress,aps,floatfix]{revtex4-1}

\usepackage{longtable}
\usepackage{graphicx}
\usepackage{dcolumn}
\usepackage{bm}
\usepackage[dvipsnames,usenames]{color}
\usepackage{subfigure}
\usepackage{multirow}
\usepackage{amssymb}
\usepackage{mdwlist}

\def\today{\space\number\day\space\ifcase\month\or January\or February\or
    March\or April\or May\or June\or July\or August\or September\or October\or
    November\or December\fi\space\number\year}

\newcommand{\iso}[2]{$^{#1}$#2}

\begin{document}

\setpagewiselinenumbers
\linenumbers

\preprint{Draft 1.0}

\title{Combining Stochastics and Analytics for a Fast Monte Carlo Decay Chain Generator}

\author{K. Kazkaz\footnote{Corresponding author, kareem@llnl.gov}}\affiliation{Lawrence Livermore National Laboratory, 7000 East Ave., Livermore CA 94551}

\author{N. Walsh}\affiliation{University of California Davis, Dept. of Physics, One Shields Ave., Davis CA 95616}

\date{\today}

\begin{abstract}
Various Monte Carlo programs, developed either by small groups or widely available, have been used to calculate the effects of decays of radioactive chains, from the original parent nucleus to the final stable isotopes. These chains include uranium, thorium, radon, and others, and generally have long-lived parent nuclei. Generating decays within these chains requires a certain amount of computing overhead related to simulating unnecessary decays, time-ordering the final results in post-processing, or both. We present a combination analytic/stochastic algorithm for creating a time-ordered set of decays with position and time correlations, and starting with an arbitrary source age. Thus the simulation costs are greatly reduced, while at the same time avoiding chronological post-processing. We discuss optimization methods within the approach to minimize calculation time.
\end{abstract}

\maketitle

%
%
\section{Introduction}
\label{s:Intro}

Various classes of physics experiments require reliable simulations of radioactive sources. Obvious examples include individual radiation detectors, but less obvious examples are low-background experiments such as are used to search for neutrinoless double-beta decay or WIMP Dark Matter. The radioactive sources are predominantly either natural background radiation that may obscure any potential signal, or calibration sources used to experimentally determine detector performance. Given the ultra-sensitive nature of current and next-generation experiments, previous understanding of detector response to bulk material radioactivity and calibration sources is usually not sufficient to meet stated scientific goals. Simulations of these detectors therefore require event generators that are highly accurate in temporal and spatial extent.

Various Monte Carlo programs are available to calculate the effects of radioactive decays in a detector setup. These programs include GEANT4~\cite{Agostinelli2003}, MCNP~\cite{Briesmeister1997}, FLUKA~\cite{Fasso2005}, or perhaps other programs written by smaller groups or individuals. Often, the radioactive decays in question derive from the decay of a single isotope to a stable nucleus, such as in the cases of \iso{137}{Cs}, \iso{60}{Co}, and \iso{40}{K}. In other cases, however, a chain of decays may occur before a stable nucleus is reached, as in the cases of \iso{238}{U}, \iso{232}{Th}, various isotopes of radon, and so on. Within these decay chains, temporal and spatial correlations are often relevant. One reason is to determine the effects of detector dead time on the acquired signal. Another reason is to interpret multiple events within any given time frame and energy range as being correlated via a single decay chain, and therefore not contributing to the signal rate.

In running Monte Carlo simulations, whether for natural backgrounds or calibration purposes, the age of the source needs to be taken into account. If the source is young and the chain has not reached secular equilibrium, the late decays will have less activity than if the chain is in secular equilibrium. In the case of background environmental radiation, the decay chains are often taken to be in secular equilibrium, and the effective source age can be on the order of millions of years. In the case of manufactured sources, the age will usually be much shorter, e.g. up to 20 years or so in the case of \iso{228}{Th}.

One simple and straightforward method to account for source age is to simulate many decays, starting with the top-most isotope, and record the time of each decay in the chain. Meanwhile, the time of each subsequent top-most decay is determined using that isotope's activity. In the analysis step, a cut can be made rejecting all events before the source age. If this approach is used, the number of decays required to achieve secular equilibrium can be very large. For example, the longest-lived descendant of \iso{238}{U} is \iso{234}{U}, with a half-life of 245.5 thousand years. Secular equilibrium is obtain after approximately 10 half-lives of the longest-lived daughter, which in this example means 2.5 million years of simulated events. If the decay activity were taken to be a mere 1 mBq, this approach still requires on the order of $10^{13}$ decays to process---an untenably large number of decays, even by modern computer standards.

It may be possible to solve this issue by incorporating an analysis of detector response time, although there are a couple disadvantages to this approach. One is that any event generator created in this manner would be applicable only for that specific kind of detector. Another is that the total source activity must also be taken into account; when simulating low-background events, the event generators may incorporate one detector response time, while simulating calibration events may require a different detector response time. The event generator would become even more specialized, and lose its general applicability.

A second difficulty comes in post-processing the data to put it in chronological order. Continuing the above example of a \iso{238}{U} source, the decay of one isotope in a source is not necessarily directly followed in time by the decay of its own daughter. Rather, the next decay might derive from a different \iso{238}{U} nucleus. For this reason, there may be effects of detector pileup or dead time that cannot be taken into account in the simulation by simply following the time structure of decay from a single parent to its eventual stable descendant. To solve this problem, the data can be post-processed so that all decays are placed in chronological order, regardless of their individual ancestor isotopes. A single simulation run can involve billions of energy depositions, and the computer resources necessary to sort these events may be substantial.

Accurate temporal correlations are, of course, not {\it always} the top priority in a detector simulation. Historically, the background levels of various materials have not always been known, and the simulation data rate simply ends up being scaled to the experimental data to infer the physical source rates. In such situations, the time of the events is usually not incorporated into the analysis. Once the absolute levels of radioactivity are known, however, more accurate simulations can be run with the correct inter-event timing.

In this paper, we present an algorithm to obtain time-ordered events from a decay chain with an arbitrary source age, while preserving temporal and spatial correlations. Our approach incorporates data structures to minimize the time required for both random search of the accumulated history of many decays, as well as insertion and deletion operations. We also present methods to reduce the recorded decay history data, minimizing computer memory usage. We evaluate our results using the \iso{232}{Th} and \iso{228}{Th} decay chains. We also benchmark the time and memory usage for various starting activities and total number of desired decays in the data set.

%
%
\section{Combining stochastic and deterministic methods for decay chain generation}
\label{s:Combination}

A purely stochastic approach to decay chain generation is intuitively compatible with Monte Carlo simulations, and its results have a high degree of reliability. In such an approach, a top-most nucleus, e.g., \iso{238}{U} for the \iso{238}{U} chain, decays, followed by a daughter. The daughter decays after a time randomly sampled from the appropriate exponential time distribution with the well-known form

\begin{equation}
\label{eq:expdist}
\mbox{Probability} \sim e^{-t\,ln(2) / T_{1/2}}
\end{equation}

\noindent
where $T_{1/2}$ is the half life of the parent. The daughter itself then decays, with the time until the next decay determined by the next daughter's half life, and so on until a stable nucleus is reached. The time between top-most decays is sampled from the distribution $e^{-t r}$, where $r$ is the activity of decays of the top-most radionuclide. This approach, while conceptually simple, often cannot be implemented because of the concerns regarding age of the source and computing requirements discussed in the previous section.

A second, more subtle problem with the purely stochastic approach is the limitations of standard data types. The double floating point type, for example, only has 16 decimal digits of accuracy. Over the course of the 2.5 million years it takes to obtain equilibrium in a \iso{238}{U} decay chain, a double floating type can maintain accuracy only to approximately 10 ms. Yet the shortest-lived daughter in the chain, \iso{214}{Po}, has a half-life of 164.3 $\mu$s. In a purely stochastic implementation, a \iso{214}{Po} decay would occur simultaneously with its parent, \iso{214}{Bi}, even though temporal separation of hundreds of microseconds is well within the capabilities of many physical detectors today. While this discrepancy between simulation and experiment may seem trivial, resolving the issue would alleviate any concerns over this artificial pileup of events.

It may be enticing to use a deterministic approach to calculate the decays from a chain. The equations involved have long been solved by H. Bateman~\cite{Bateman1910}. Unfortunately, a purely analytic approach is incompatible with a Monte Carlo simulation of radioactivity, as the individual decays lose their spatial correlations. Consider the case of a \iso{212}{Bi} nucleus decaying into a \iso{212}{Po} nucleus. Because \iso{212}{Po} has a half life of 300 ns, its own decay will be tightly correlated to the position of the \iso{212}{Bi}, and thus information from each decay must be passed along to progeny nuclei. This information is lost in the purely deterministic approach.

We have therefore combined the two approaches in the current treatment. We employ Bateman's equations to calculate populations of all isotopes in the decay chain at any given source age. With those populations known, we proceed using a stochastic approach. In this way, we avoid the severe computational overhead while preserving the spatial correlations within a decay chain.

\subsection{The algorithm}
\label{ss:Basics}

The inputs to the algorithm for any given decay chain, apart from defining the chain itself, are the starting activity of the top-most isotope, the age of the source, and the number of events desired in the resulting decay record. Using these inputs makes direct comparison to known sources more straightforward, as commercial radioactive sources have an activity and manufacturing time stamp. In the case of environmental radiation, such as background \iso{238}{U} or \iso{232}{Th}, the steady-state activity is usually measured {\it in situ}, and the age of the source then set to 10 times the half life of the longest-lived daughter nucleus to reach secular equilibrium.

The first step in the process is to calculate the starting populations of all isotopes in the chain using the Bateman equations. In our own implementation, we used the Moral and Pacheco approach to the Bateman equations~\cite{Moral2003}. Following Moral and Pacheco's recommendation, we used {\it Mathematica}~\cite{Mathematica8} to solve the equations, and entered the solutions into our decay generator computer code. We assumed, as in~\cite{Moral2003}, that the population of the chain before source aging was entirely in the top-most isotope.

Because the population equations are analytic, it is possible to obtain a fractional population. To limit calculations performed on unphysical numbers, if the population ever fell below 0.5 in our treatment, we adjusted the population to identically 0. Once the populations were determined, the starting activity of the $i^{th}$ isotope, $r_i$, was calculated using the equation

\begin{equation}
\label{eq:activity}
r_i = -\frac{ln(2)}{T_{1/2}} N_i
\end{equation}

\noindent
where $N_i$ is the population of the $i^{th}$ isotope at any given time.

Within the current algorithm, we distinguished between decays from the primary population, i.e., the population after source aging, and decays from the subsequent chains. Additionally, we distinguished and tracked the time of primary decays (``primary time'') and the time of chain decays (``chain time''). The algorithm is as follows:

\begin{enumerate}
\setlength{\itemsep}{1pt}
\setlength{\parskip}{0pt}
\setlength{\parsep}{0pt}
\item For the calculated surviving primary populations, determine rates for each isotope using Eq.~\eqref{eq:activity} 
\item Increment the primary time based on an exponential random sampling, using the total surviving primary decay activity in Eq.~\eqref{eq:expdist}
\item Randomly select the next primary decay isotope, weighted by the activities of the surviving primary populations
\item Reduce the primary population of that isotope by 1
\item Set the chain time to the primary time
\item Generate a location for the new primary decay
\item Add the decay to the record history, including isotope, position, and time
\item If the isotope has a radioactive daughter, increment the chain time by a random number distributed according to Eq.~\eqref{eq:expdist}, with the appropriate $T_{1/2}$ of the isotope, and propagate the decay location to the new decay
\item If the number of decays in the data record is greater than the number requested, remove the last entry from the record
\item Continue the previous three steps until a stable isotope is reached
\item Return to the first step unless the requested number of primary decays has been reached
\end{enumerate}

As can be seen from this algorithm, the correct spatial and temporal correlations through an entire decay chain, starting at an arbitrary source age, is maintained. Additionally, the issue of temporal resolution is largely resolved. Consider the case of accumulating 10 million decay events in the data record. If the total activity were a constant 1 Becquerel, then the last decays in the data set would have a time resolution on the order of 1 ns.

Some care must be taken in determining the correct termination conditions. If 10 million events are desired in the decay record, it is incorrect to simply stop the calculation after 10 million events have been entered. \iso{232}{Th}, for example, has 10 products in its decay chain, so 1 million decays of \iso{232}{Th} would provide the desired number of events. Unfortunately, after those 1 million decays of \iso{232}{Th}, with a relatively steady activity, the event rate will drop precipitously, as the only decays after that point in time depend on the 5.75-year half life of \iso{228}{Ra}. To avoid this unphysical data record, 10 million full decays of the starting population should be performed.

The termination conditions can also be optimized. For example, if the number of requested decays has been entered into the data record, and the decay time is beyond the time of the last decay in the record, there is no point in continuing that particular chain. Other optimizations may also exist, depending on the specific circumstances of starting activities, number of requested events, and so on.

As a comment on the algorithm, it may be more natural when running a simulation to require a set amount of simulated time, rather than a set number of simulated events. Unfortunately, this introduces a kind of tension with, for example, GEANT4, where the number of events is specified as the input. To run for a specific length of time, we recommend calculating beforehand how many events there should be in the desired data set, and augmenting that number appropriately to ensure that the full desired simulation time is accumulated in the event record.

\subsection{Data structure}
\label{ss:DataStructure}

Our approach required an appropriate data structure to maintain and time-order the events as they were generated by the algorithm. The standard C++ sequence container classes of {\it vectors} or {\it lists} are disadvantageous because of computing overhead. Vectors have the advantage of random access, allowing for a find operation with O(log n) overhead. Unfortunately, insertion and deletion operations are computationally expensive, with O(n) requirements. Conversely, lists handle insertion and deletion in O(1) time, but without random access, finding the correct location to place a decay is an O(n) operation. We therefore decided to use a binary search tree data structure, where a find operation is performed in O(log n) time, while insertion and deletion are still maintained at O(1), thus combining the advantages of both vectors and lists. The C++ associative container class {\it multimap} would also be appropriate, although we implemented our own binary search tree to avoid residual overhead of a multimap.

Each subsequent primary decay occurs later in time than the one before it. Similarly, a chain decay also occurs after its parent decay. These are obvious physical facts, but can negatively affect the performance of the data structure. Because each decay time occurs after the one before it, it is possible to end up with a semi-degenerate binary search tree, far more heavily weighted to one side than the other. If this were to occur, find operations approach O(n), and we lose the advantage of the search tree.

\begin{figure}[h!!!]
\centering
\includegraphics[width=8.5cm]{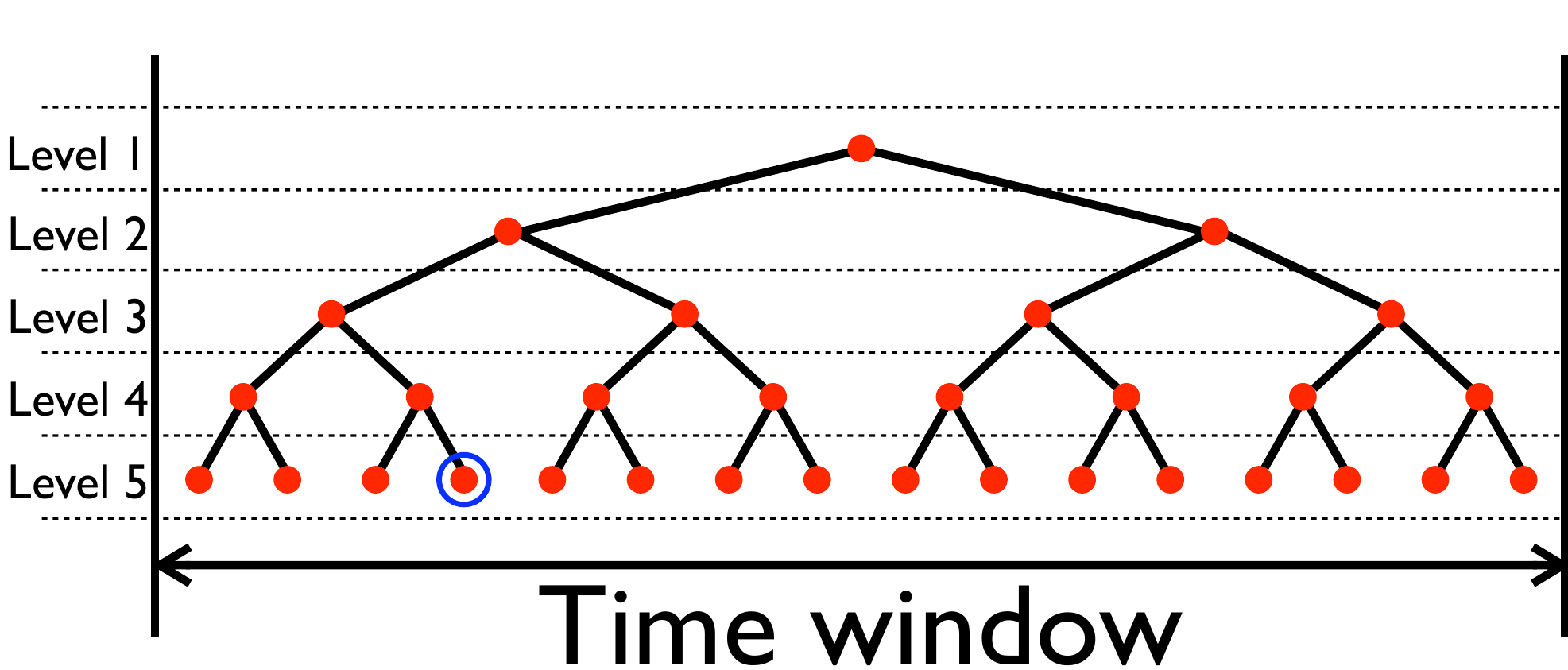}
\parbox{8.5cm}{\vspace{5pt}\caption{\small{The structure of empty nodes in the pre-seeded binary search tree. By subdividing the time window, degeneracy in the data structure can be minimized. The nodes were entered in the order of their level number, and split the time between the two nodes closest in time, e.g., the time of the node circled in blue is at the average of its parent node and the left node of Level 2.}}
\label{fig:EmptyNodeStructure}}
\end{figure}

To fix this situation, we pre-seeded the search tree with empty nodes to install an evenly-spaced structure, and thus avoid a quasi-degenerate tree. We calculated the anticipated simulated time window to acquire all requested decays. This time window was approximate, as the ever-reducing primary population altered the time of the last decay in a non-trivial way. We then sub-divided this time window into a number of levels, where each level contained twice as many empty nodes as the level above it. The spacing of each level was such that the node split in half the time to either side of the upper nodes (see Fig.~\ref{fig:EmptyNodeStructure}). The levels also had to be inserted in the order of their level, again to avoid quasi-degeneracy.

The time savings of establishing an empty node structure in the binary search tree varied with the number of events already entered. If only one level was used, the time it took to populate the tree with $10^5$ decays took approximately ten times as long as if 20 levels were used. As the tree is populated, however, it becomes increasingly balanced, since primary decays tend to occur at earlier times and chain decays at later times. This leads to a counter-intuitive result that the time savings afforded by pre-seeding the tree diminish with an increasing number of event. The computing time required to populate a tree with $10^7$ events did not change appreciably with the number of pre-seeded levels. The authors therefore recommend using 20 pre-seed levels, or approximately 1 million empty nodes, to provide adequate protection against degeneracy, at the expense of less than 2 seconds of computing time and 40 megabytes of memory. If a large number of independent simulations, each with $10^5$ decays, is required, the time savings might be substantial.

%
%
\section{Application and evaluation}
\label{s:AppAndEval}

We evaluated our approach using the \iso{232}{Th} and \iso{228}{Th} decay chains. The former introduces a case of a near-constant decay activity in the top-most isotope, while in the latter case the population of the parent nucleus noticeably decays with time.

\subsection{Using a \iso{232}{Th} source}
\label{ss:Th232}

\begin{figure}[t]
\centering
\subfigure{\includegraphics[width=8.5cm] {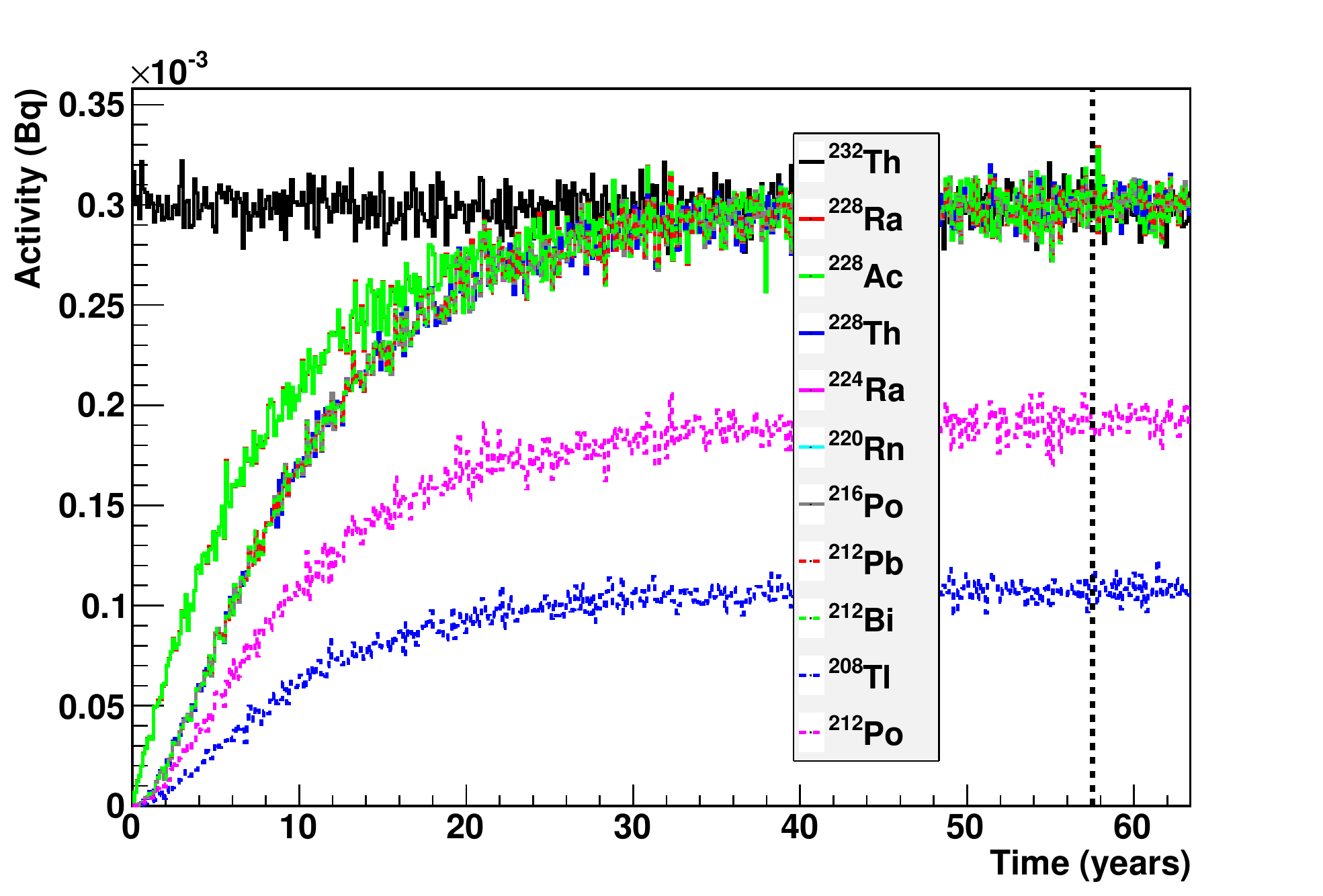}}
\subfigure{\includegraphics[width=8.5cm] {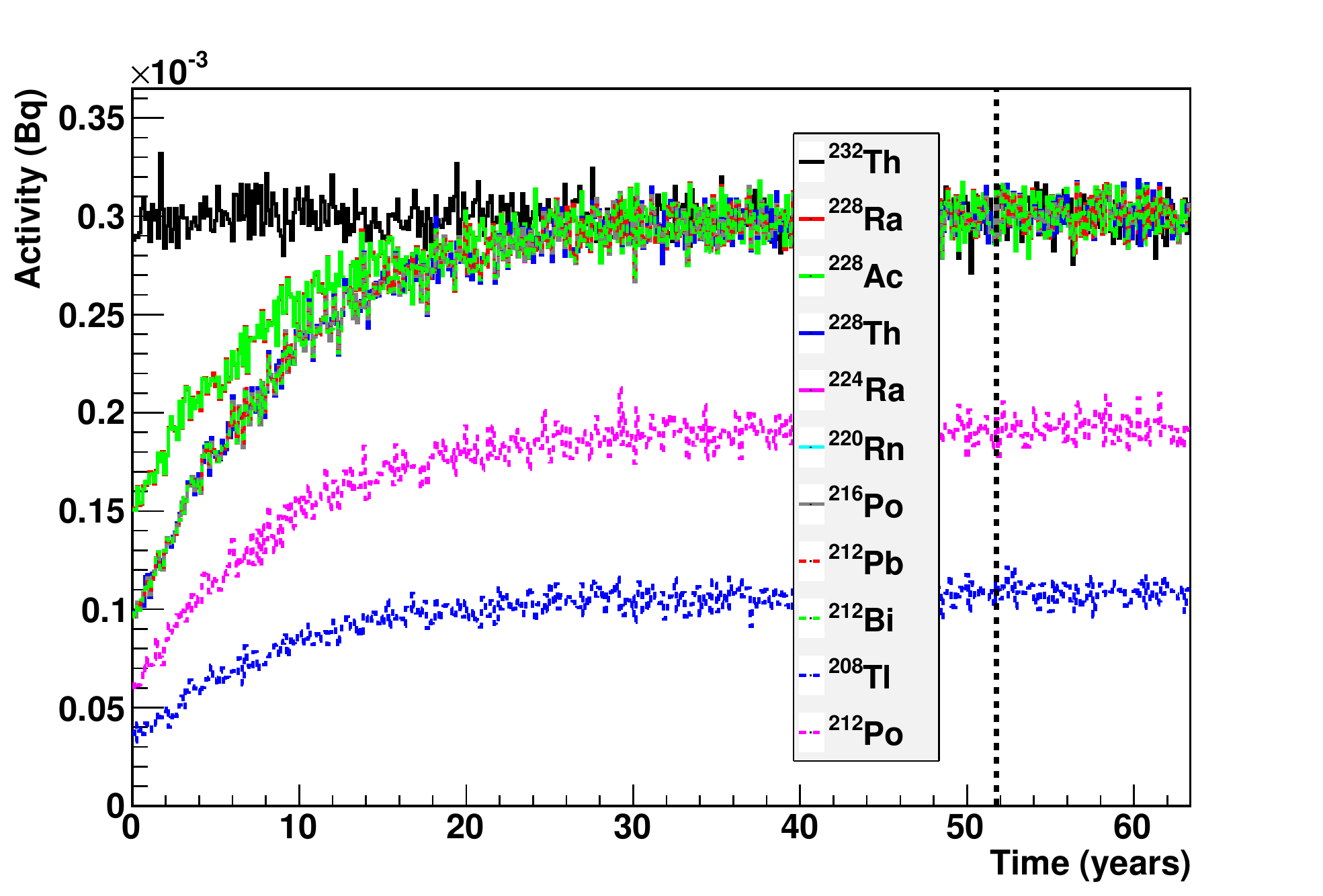}}
\subfigure{\includegraphics[width=8.5cm] {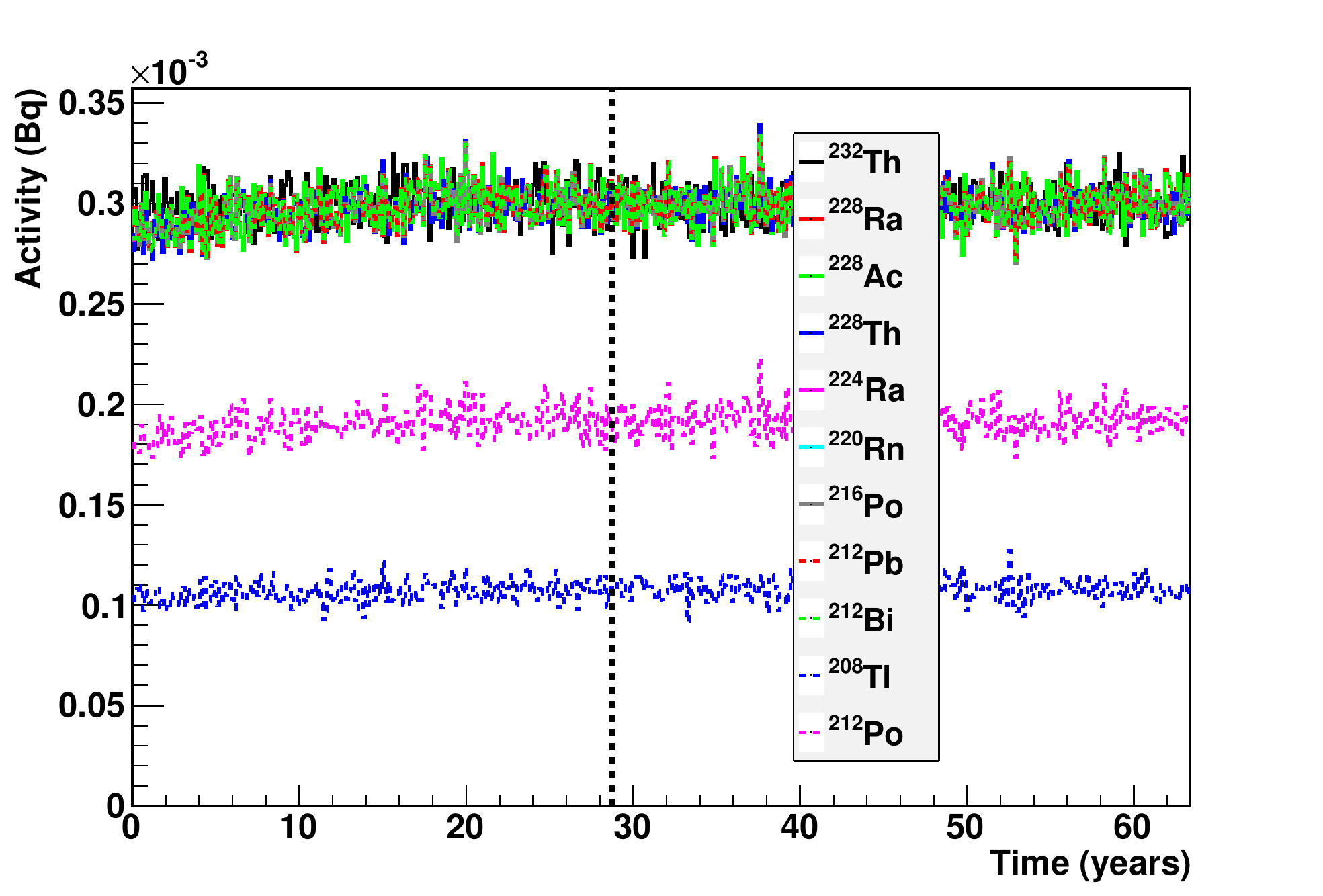}}
\parbox{8.5cm}{\vspace{5pt}\caption{\small{Decay activities of isotopes within the \iso{232}{Th} decay chain. The top figure has a source age of 0 years, the middle a source age of 1 half life of \iso{228}{Ra}, and the bottom figure a source age of 5 half lives of \iso{228}{Ra}. In each case, the vertical dashed line marks 10 half lives from the creation of the source. For reference, the half life of \iso{228}{Ra} is 5.75 years.}}
\label{fig:232ThRates}}
\end{figure}

Figure~\ref{fig:232ThRates} show the decay activities of all the isotopes in the \iso{232}{Th} decay chain. For the three figures, the source age was set to 0, 1, and 5 times the half life of the longest-lived daughter in the chain, \iso{228}{Ra}. The figures demonstrate the reliability of combining the analytic Bateman calculations to determine the starting population, followed by stochastic decays within the chain. For these calculations, the starting decay activity of the \iso{232}{Th} was kept constant at 0.3 mBq, and 10 million decays were recorded.

Figure~\ref{fig:232ThRates} also shows how the current approach also properly handles branching within the decay chain. In this case, \iso{212}{Bi} decays to \iso{208}{Tl} (branching ratio = 35.93\%), \iso{212}{Po} (branching ratio = 64.05\%), and directly to \iso{208}{Pb} (branching ratio = 0.023\%). The rates of \iso{208}{Tl} and \iso{212}{Po} are therefore lower than the others.

\subsection{Using a \iso{228}{Th} source}
\label{ss:Th228}

\begin{figure}[t]
\centering
\subfigure{\includegraphics[width=8.5cm] {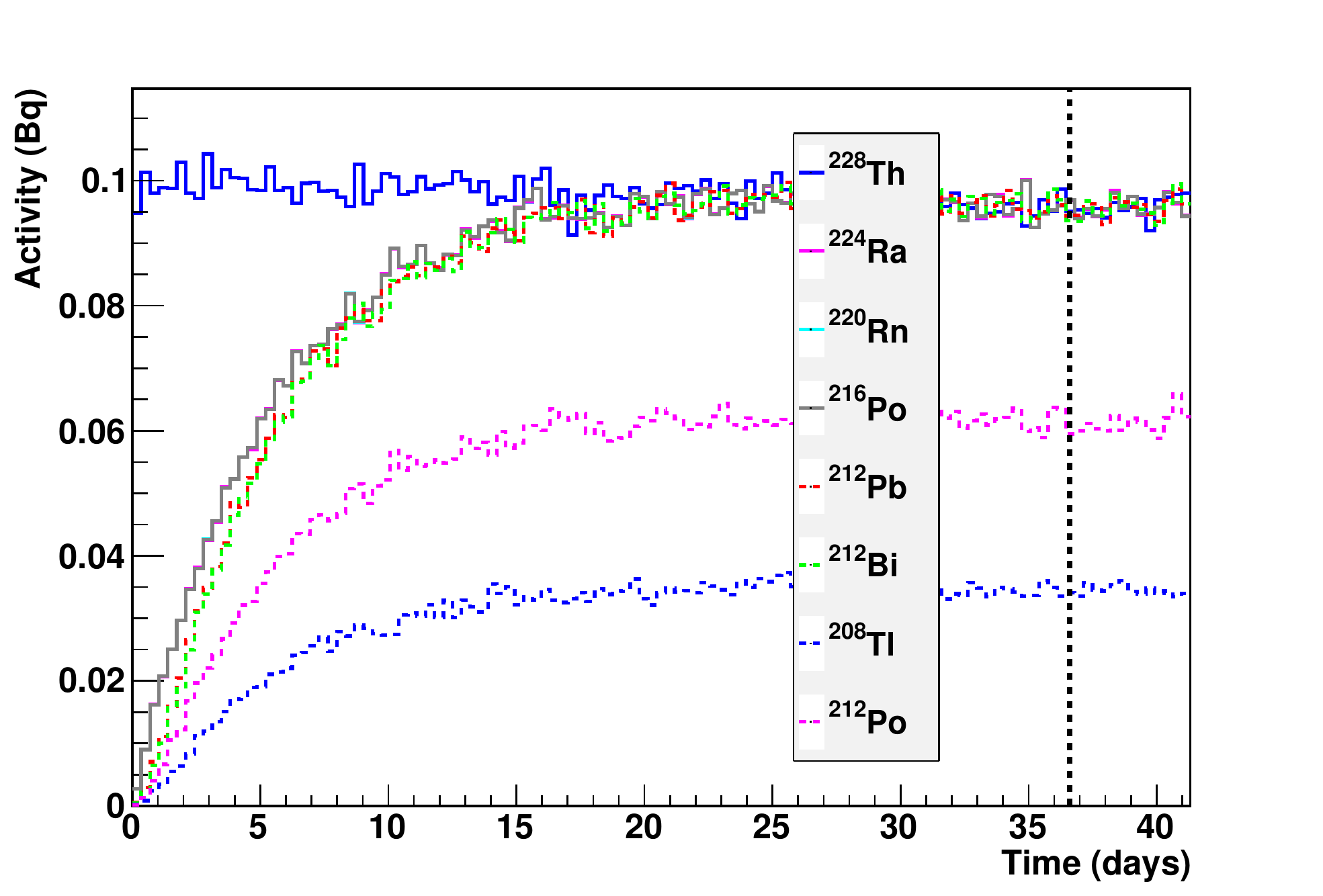}}
\subfigure{\includegraphics[width=8.5cm] {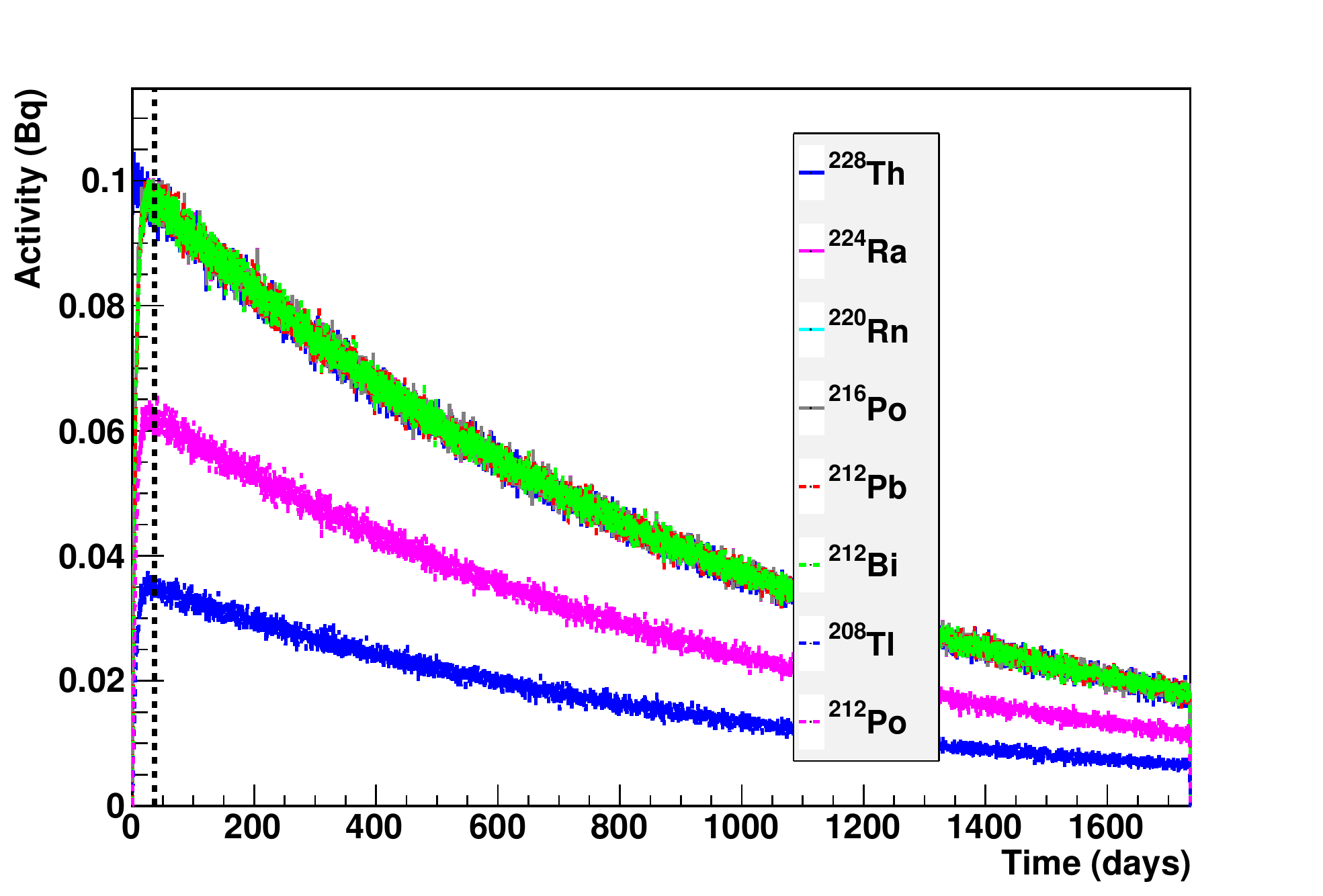}}
\parbox{8.5cm}{\vspace{5pt}\caption{\small{Decay activities of isotopes within the \iso{228}{Th} decay chain. The initial build-up of daughter isotopes is visible in the top plot, while the bottom plot shows the long-term behavior of the decays. The vertical dashed line marks 10 times the half life of \iso{224}{Ra}. For reference, the half life of \iso{224}{Ra} is 3.6 days.}}
\label{fig:228ThRates}}
\end{figure}

Figure~\ref{fig:228ThRates} demonstrates the validity of our approach when the top-most isotope's half life is short relative to the time window. The daughter isotopes are clearly populated from the starting parent population. The exponential trend in the overall decay rate, dictated by the 1.9-year half life of \iso{228}{Th}, is also clear. For these calculations, a starting decay activity of 0.1 Bq was used, and 50 million decays were recorded.

\subsection{Benchmarking}
\label{ss:Benchmarking}

We benchmarked the performance of our decay generator. Table~\ref{tab:Benchmarking} shows the time required to perform the requested calculations. The calculations were run on a desktop with a 2.26 GHz Intel E5520 Xeon chip and 8 GB of RAM, running MacOSX 10.5.8. The vast majority of the memory usage was taken up by the binary search tree data structure, where each node of which required 30 bytes of RAM. 20 levels of pre-seeded nodes required approximately 1 million entries, and the rest of the entries were specified at run time. Thus a data record of 50 million events required approximately 1.5 GB of memory. This is a large amount, certainly, but still within the capabilities of any modern computer.

Accurate extrapolations of the computing time required to create the event record are not straightforward, as they involve a non-trivial interplay between the pre-seeded binary search tree, the rate of primary decays, and the required number of traversals of a full decay chain. Broadly speaking, however, the computing time scales with the number of required events, and the longest computing time occurs when the source age is 0.

\begin{table}[t!!!]
\renewcommand*\arraystretch{1.2}
\caption{\small{Benchmarking the decay chain generator. The second row in the headings contains the number of decays in the final data record. The third row in the headings is the source age, where the value of $T_{1/2}^a$ is 5.75 years, and 3.6 days for $T_{1/2}^b$. All entries below the headings are in seconds, except for the column on the left. The first row of entries for the \iso{228}{Th} decays is absent because the starting activity implies a starting population too low to provide the full set of decays. Each entry in the table represents the average and standard deviation from 10 trials. See text for details.}}
\begin{center}
\begin{tabular}{|c|c|c|c|c|}
\hline
\multirow{3}{*}{$\begin{array}{c} \mbox{Starting} \\ \mbox{activity} \\ \mbox{(Bq)} \end{array}$}
& \multicolumn{4}{c|}{\iso{232}{Th} Decay Chain} \\
& \multicolumn{2}{c|}{$10^7$} & \multicolumn{2}{c|}{$5 \times 10^7$} \\
& 0 $T_{1/2}^a$ & 5 $T_{1/2}^a$ & 0 $T_{1/2}^a$ & 5 $T_{1/2}^a$ \\
\hline
$10^{-6}$	& $9.2 \pm 0.4$		& $9.7 \pm 0.7$	& $48.9 \pm 1.2$	& $49.9 \pm 1.0$	\\
1		& $93.9 \pm 1.6$		& $48.2 \pm 1.5$	& $522 \pm 14$	& $263 \pm 11$	\\
$10^6$	& $97.7 \pm 1.8$		& $75.7 \pm 1.7$	& $616 \pm 17$	& $423 \pm 18$	\\
\hline
\hline
\multirow{3}{*}{$\begin{array}{c} \mbox{Starting} \\ \mbox{activity} \\ \mbox{(Bq)} \end{array}$}
& \multicolumn{4}{c|}{\iso{228}{Th} Decay Chain} \\
& \multicolumn{2}{c|}{$10^7$} & \multicolumn{2}{c|}{$5 \times 10^7$} \\
& 0 $T_{1/2}^b$ & 5 $T_{1/2}^b$ & 0 $T_{1/2}^b$ & 5 $T_{1/2}^b$ \\
\hline
$10^{-6}$	& N/A	& N/A	& N/A 		& N/A \\
1		& $19.3 \pm 0.7$		& $19.9 \pm 0.9$		& $93.8 \pm 1.1$		& $91.9 \pm 1.3$ \\
$10^6$	& $81.2 \pm 1.5$		& $56.1 \pm 1.3$		& $520 \pm 15$		& $320 \pm 13$ \\
\hline
\end{tabular}
\end{center}
\label{tab:Benchmarking}
\end{table}

\section{Conclusions}
\label{s:Conclusions}

The approach described in this paper for decay chain generation can be used to calculate a full, time-ordered history of decays from a radioactive source with an extended decay chain, while preserving both time and position correlations between decays within a single traversal of the chain. Once these calculations are performed, the data record can be passed to a Monte Carlo program to produce the decay products themselves at the recorded times and positions. Such an event generator is useful for both basic science simulations as well as detector component evaluation, and does not make any assumptions regarding detector response time or data acquisition performance. As such, it is appropriate for a wide variety of applications.

Using a purely stochastic or a purely analytic approach to decay chain generation both contain strong disadvantages. The two approaches can be combined, however, to great effect, with the analytics used to generate starting populations among the decay nuclei, and stochastics used to propagate temporal and spatial correlations. Using the combination, a large number of time-ordered decays can be generated in just a few minutes, with the computational overhead being driven primarily by the requested number of decays in the record.

The data structure of the record must be chosen to minimize both search and insertion/deletion operation overhead, and thus the authors recommend a binary search tree. To minimize degeneracy, the search tree can be pre-seeded with empty nodes to apply a structure within which the decays can be recorded.

The authors would like to thank Jason Detwiler for reviewing the manuscript, and Aric Stewart for recommending  binary search trees for this application. Prepared by LLNL under Contract DE-AC52-07NA27344. Funded by Lab-wide LDRD. LLNL-JRNL-478911.


%
%
\bibliography{apssamp}

\end{document}